\documentclass{phostproc}

\title{Combining multiple structural inversions to constrain the Solar modelling problem}
\author{G. Buldgen,$^{1,4}$
S.J.A.J. Salmon,$^{2}$
A. Noels,$^{2}$
V.A. Baturin,$^{3}$
P. Eggenberger,$^{4}$
G. Meynet,$^{4}$
A. Miglio,$^{1}$}

\affiliation{$^{1}$ School of Physics and Astronomy, University of Birmingham, Edgbaston, Birmingham B15 2TT, UK.\\ 
$^{2}$ Institut d'Astrophysique et Géophysique, University of Liège, Liège, Belgium.\\
$^{3}$ Sternberg Astronomical Institute, Lomonosov Moscow State University, 119234 Moscow, Russia.\\
$^{4}$ Observatoire de Genève, Université de Genève, 51 Ch. Des Maillettes, CH-1290 Sauverny, Suisse.}

\shorttitle{Combining multiple structural inversions
to constrain the Solar modelling problem}
\shortauthors{G. Buldgen \textit{et al.}}

\abs{The Sun is the most studied of stars and a laboratory of fundamental physics. However, the understanding of our star is stained by the solar modelling problem which can stem from various causes. We combine inversions of sound speed, an entropy proxy and the Ledoux discriminant with the position of the base of the convective zone and the photospheric helium abundance to test combinations of ingredients such as equation of state, abundance and opacity tables. We study the potential of the inversions to constrain ad-hoc opacity modifications and additional mixing in the Sun. We show that they provide constraints on these modifications to the ingredients and that the solar problem likely occurs from various sources and using phase shifts with our approach is the next step to take.}

\begin{document}

\maketitle

\section{Introduction}

The Sun is the most constrained and well-studied of all stars and a laboratory of fundamental physics. Currently, the physical ingredients entering solar models are used as a reference to study all other stars observed in the Universe.

However, our understanding of the Sun is imperfect, as illustrated by the current debate on the solar heavy element abundances. This problem is intimately related to the microscopic and macroscopic physical ingredients used to compute solar models. The uncertainties on these physical processes are the main limitations when determining stellar fundamental parameters, such as mass, radius and ages, with asteroseismic data from CoRoT and \textit{Kepler}, or future observations from TESS and Plato. 

In this study, we combine sound speed, entropy proxy and the Ledoux discriminant inversions to test ingredients such as equation of state, abundance and opacity tables. We study the potential of inversions to constrain ad-hoc opacity modifications and additional mixing in the Sun. We show that they provide constraints and that the solar problem likely occurs from various sources and using phase shifts with our approach is the next step to take.

\section{Testing standard ingredients}\label{secSTD}

In this section, we test various ingredients of standard solar models and analyze their influence on sound speed inversions. The models have been computed using the Liège stellar evolution code (CLES, \citet{ScuflaireCles}), their oscillation frequencies have been determined using the Liège stellar oscillation code (LOSC, \citet{ScuflaireOsc}) and the inversions have been computed using the InversionKit software and the SOLA inversion technique \citep{Pijpers}, following the guidelines of \citet{RabelloParam} for the optimization of the trade-off parameters. The inversion equations are the classical linear relations derived by \citet{Dziemboswki90} using the variational principle of adiabatic stellar oscillations (see e.g. \citet{LyndenBell}). We computed inversions for the squared adiabatic sound speed, $c^{2}=\frac{\Gamma_{1}P}{\rho}$, with $P$ the pressure, $\rho$ the density and $\Gamma_{1}$ the adiabatic exponent defined as $\frac{\partial \ln P}{\partial \ln \rho} \vert_{S}$, with $S$ the entropy of the stellar plasma as well as for an entropy proxy, defined as $S_{5/3}=\frac{P}{\rho^{5/3}}$ and the Ledoux discriminant, defined as $A=\frac{d \ln P}{d \ln r}-\frac{1}{\Gamma_{1}}\frac{d \ln \rho}{d \ln r}$.

Here, we limit ourselves to present the comparisons for models using various abundance and opacity tables, as these have the largest impact on the inversion results. We refer the reader to \citet{BULDGEN} for additional information. For our comparisons, we chose to keep one set of ingredients as reference and plot all the other inverted profiles in figures including this specific reference in order to see directly the effects of various ingredients. We used a model built using the AGSS$09$ abundances \citep{AGSS09}, the FreeEOS equation of state \citep{Irwin}, the opal opacities \citep{OPAL}, the mixing-length theory of convection \citep{Cox}, the formalism for microscopic diffusion by \citet{Thoul}, and the nuclear reaction rates from \citet{Adelberger}. All models also include effects of conduction from \citet{Potekhin} and from \citet{Cassisi}, as well as low-temperature opacities from \citet{Ferguson}. We used grey atmosphere models in the Eddington approximation in all our models. 

\begin{figure*}
	\centering
	\includegraphics[width=0.75\linewidth]{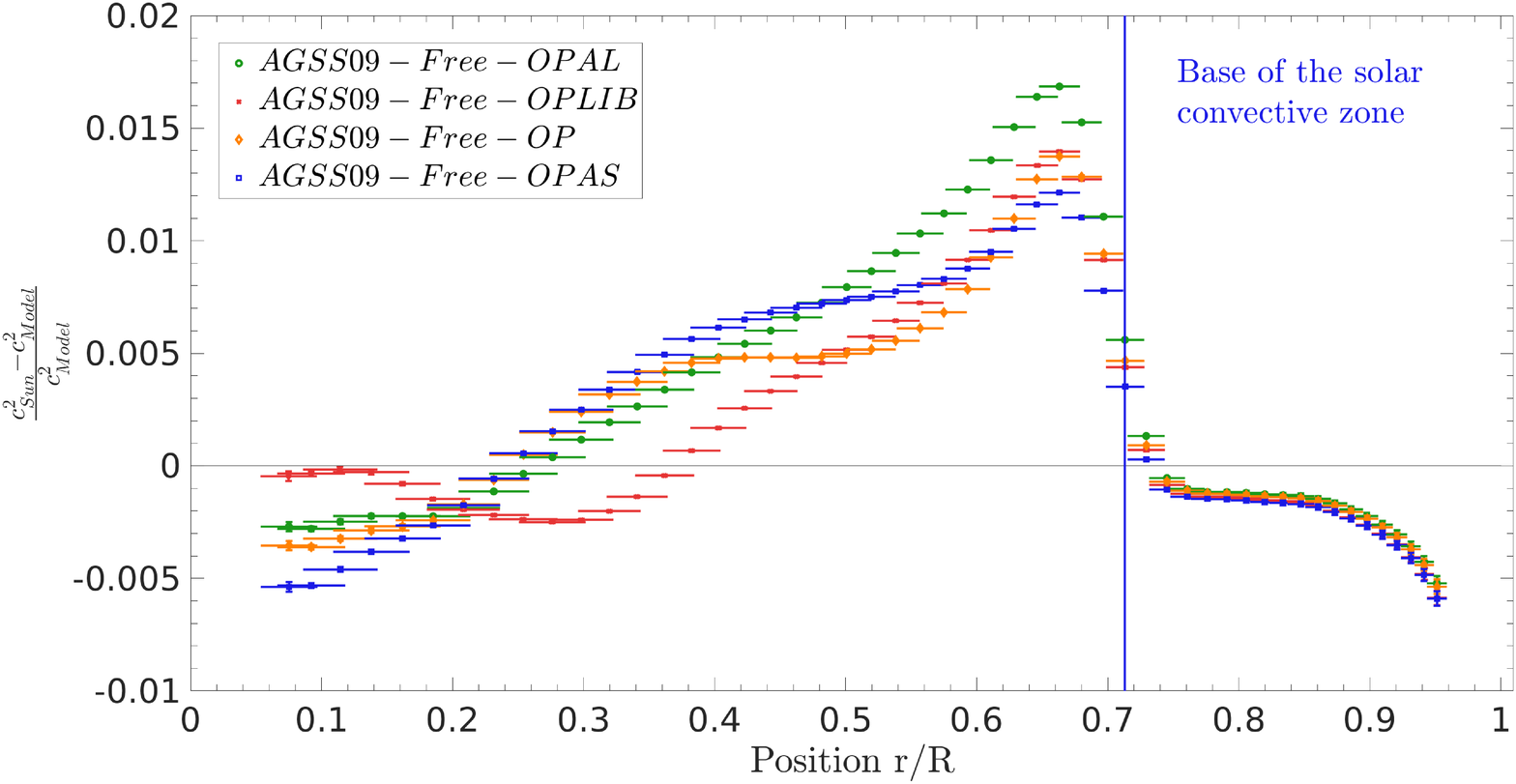}
	\caption{Relative differences in squared adiabatic sound speed between the Sun and standard solar models built using various opacity tables.}
	\label{figC2STDOpac}
\end{figure*}

\begin{figure*}
	\centering
	\includegraphics[width=0.75\linewidth]{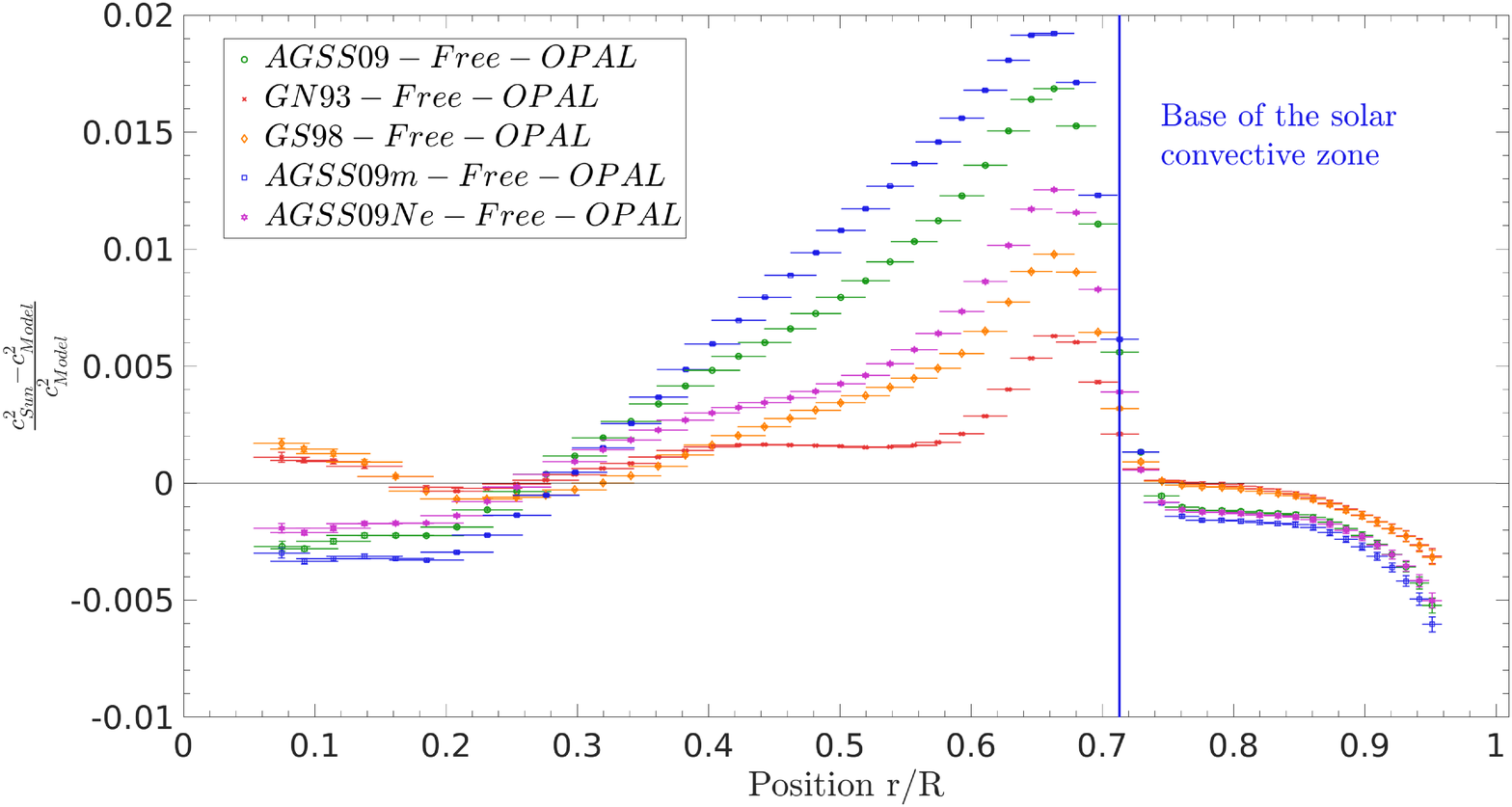}
	\caption{Relative differences in squared adiabatic sound speed between the Sun and standard solar models built using various abundance tables of the heavy elements.}
	\label{figC2STDAbund}
\end{figure*}

To test the importance of the opacity tables, we built models using the OPAL, OPAS \citep{Mondet}, OPLIB \citep{Colgan}, and OP \citep{Badnell} opacity tables. For the abundances, we used models built with the former GN$93$ and GS$98$ abundances \citep{GrevNoels,GreSauv} and models built with the more recent AGSS$09$ abundances. We also computed one table for which the abundances of C, N, O, Ne and Ar were changed to the meteoritic values, as was done in \citet{SerenelliComp}, denoted AGSS$09$m and one for which the recently suggested $40\%$ neon abundance increase was taken into account \citep{Landi,Young}, denoted AGSS$09$Ne. For each of these composition tables, the solar $Z/X$ ratio to be reproduced was adapted accordingly and opacity tables were recomputed for each abundance table. The results of these comparisons are illustrated in Figures \ref{figC2STDOpac} and \ref{figC2STDAbund} for the effects of opacity tables and chemical abundances, respectively.

From Figure \ref{figC2STDOpac}, we can see that using the more recent opacity tables (namely the OPAS and OPLIB tables) leads to a small improvement of the agreement of AGSS09 models with helioseismic data. However, this improvement is mitigated by the reduction in helium abundance in the convective zone observed in models built using these tables. This is a consequence of the reduction of opacity in most of the solar radiative zone observed in these tables, which leads to an increase in hydrogen abundance when calibrating solar models (see \citet{BULDGEN} for a discussion). Thus, it appears that the stalemate remains even if more recent tables are used.

We see in Figure \ref{figC2STDAbund} that the best agreement in therms of sound speed inversions is unsurprisingly found for standard solar models built with GS98 or GN93 abundance tables. However, we can also see that using the meteoritic abundances for refractory elements leads to a slightly larger disagreement for the AGSS09 models whereas taking into account the neon increase suggested by \citet{Landi} and \citet{Young} leads to a significant improvement of the agreement. This is again not really surprising, as neon is the third contributor to opacity at the base of the convective envelope \citep{BlancardOpacDetail}. An increase of the neon abundance, although much larger, was already suggested by \citet{Antia05}, \citet{Zaatri2007}, and \citet{Basu08} to restore the agreement between standard solar models and seismic constraints.

\section{Opacity modifications}\label{secOPAC}

\begin{figure}
	\centering
	\includegraphics[width=0.95\linewidth]{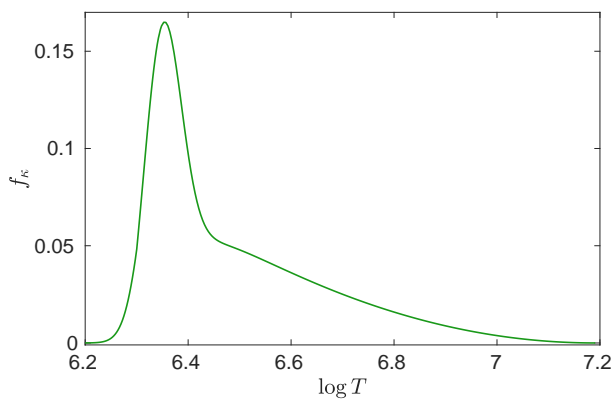}
	\caption{Modification applied to the mean Rosseland opacity of the models.}
	\label{figOpac}
\end{figure}

In section \ref{secSTD}, we have presented some inversion results using various key physical ingredients of standard solar models and shown how they affected the solar modelling problem. From these tests, we concluded that no combination of physical ingredients available was sufficient to break the current stalemate. In this section, we take a step further by analyzing the impact of ad-hoc opacity modifications on standard solar models. Our approach to modifying the mean Rosseland opacity is based on the current debate in the opacity community following the experimental determination of iron opacity in the conditions of the solar convective envelope and on private communications with Prof. A. Pradhan. From these discussions, we computed a modification of the mean Rosseland opacity using a peaked Gaussian near the base of the convective envelope followed by a steep polynomial decrease. This profiles is supposed to qualitatively reproduce the fact that the inaccuracies of opacity tables are expected to be smaller at higher temperatures. The opacity modification is made as follows:
\begin{equation}
\kappa^{'}=\kappa \left( 1 + f_{\kappa} \right),
\end{equation}
with $\kappa^{'}$ the modified opacity used for the calibration, $\kappa$ the opacity value in the table and $f_{\kappa}$ the opacity modification, illustrated in Figure \ref{figOpac}.

\begin{figure*}
	\centering
	\includegraphics[width=0.75\linewidth]{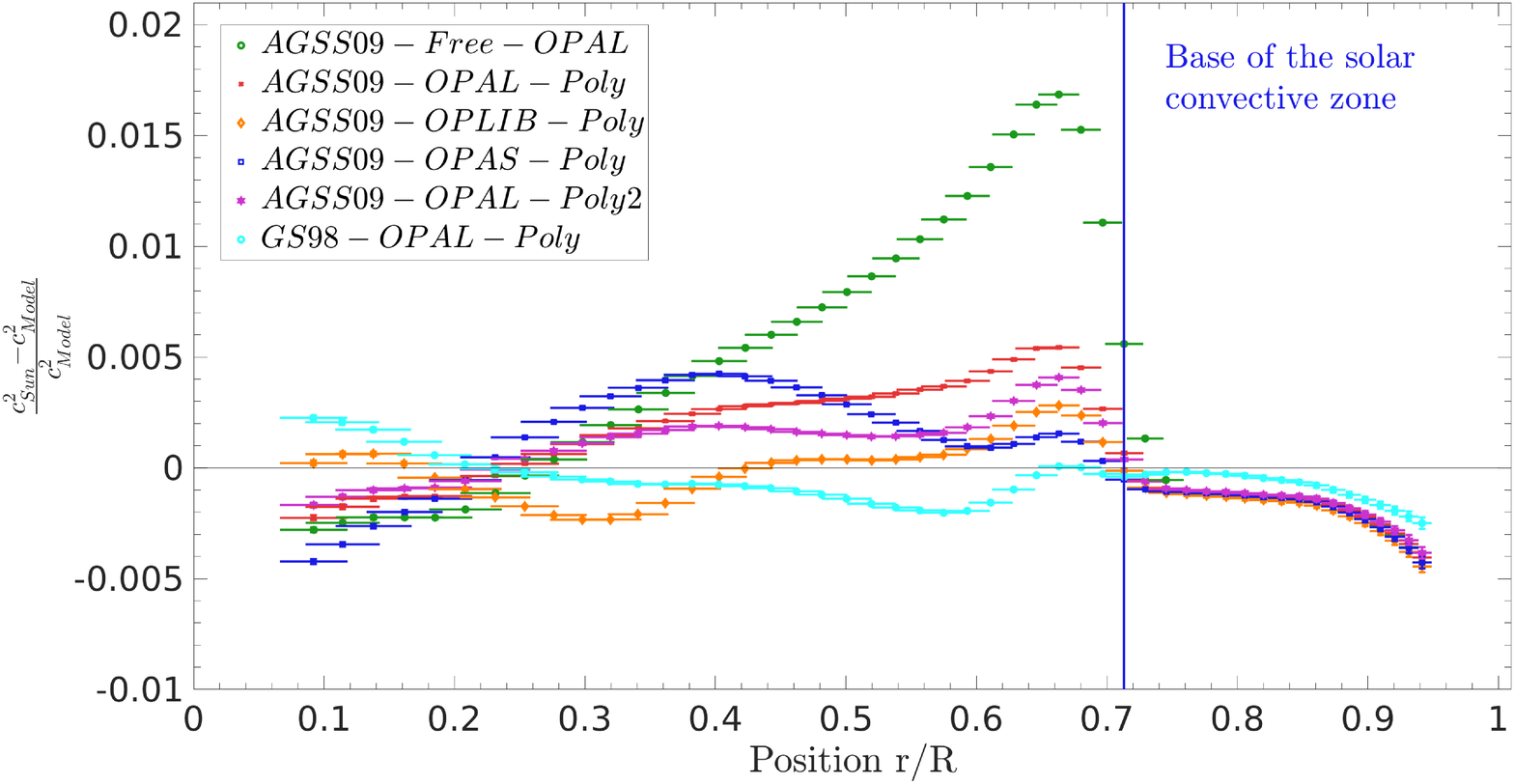}
	\caption{Relative differences in squared adiabatic sound speed between the Sun and solar models built using an ad-hoc modification of the mean Rosseland opacity.}
	\label{figC2Poly}
\end{figure*}

The impact of this opacity modification on the squared adiabatic sound speed profile of standard solar models is illustrated in Figure \ref{figC2Poly}. As we can see, a significant improvement is achieved with all standard opacity tables. However, we can see that rather large discrepancies in the deep radiative layers remain for the model built with the OPAS opacities. We also tested our opacity modification on a model built with the GS98 abundance tables. This test has demonstrated that a further increase in opacity would lead to a significant disagreement of GS98 models with helioseismic constraints, especially a too deep position of the base of the convective envelope. However, this also appeared to be the case with AGSS09 models built using the OPAS and OPLIB opacity tables. We also tested a slightly higher increase in opacity, denoted ``Poly2'' in Figure \ref{figC2Poly} on a standard model built using the OPAL opacities. In this specific case, the opacity modification induced a slightly to deep position of the base of the convective envelope. 

From Figure \ref{figC2Poly}, one could think that the agreement of these models is excellent. However, in \citet{BULDGEN}, we presented additional inversions of an entropy proxy and the Ledoux discriminant that showed the compensation effects that could be hiding behind an excellent agreement in sound speed. Moreover, none of these models presented a helium abundance in the convective envelope in agreement with the recent determinations of \citet{Vorontsov13} as well as the results of \citet{BasuYSun}. Some example of these compensation effects will be illustrated in the following section.

\section{Modified solar models}

From additional inversions presented in \citet{BULDGEN}, we could show that a good agreement in sound speed could hide some compensation effects and that using supplementary inversions could lift some of these degeneracies. Thus, we aimed for an additional improvement of the agreement of our AGSS09 models by adding a supplementary chemical mixing at the base of the convective zone, using the modification of the mean Rosseland opacity previously presented and the increase in neon abundance prescribed by \citet{Landi}, and \citet{Young}. All the models presented in this section were built using the SAHA-S equation of state \citep{Gryaznov04,Baturin}. Again, the results presented here remain qualitative and a more detailed study of the extra-mixing from both a seismic and theoretical point of view should be performed. Multiple studies have discussed the issue of properly modelling the base of the solar convective zone, which is also the seat of the solar tachocline, supposed to harbour various macroscopic processes not included in standard solar models.

\begin{figure*}
	\centering
	\includegraphics[width=0.75\linewidth]{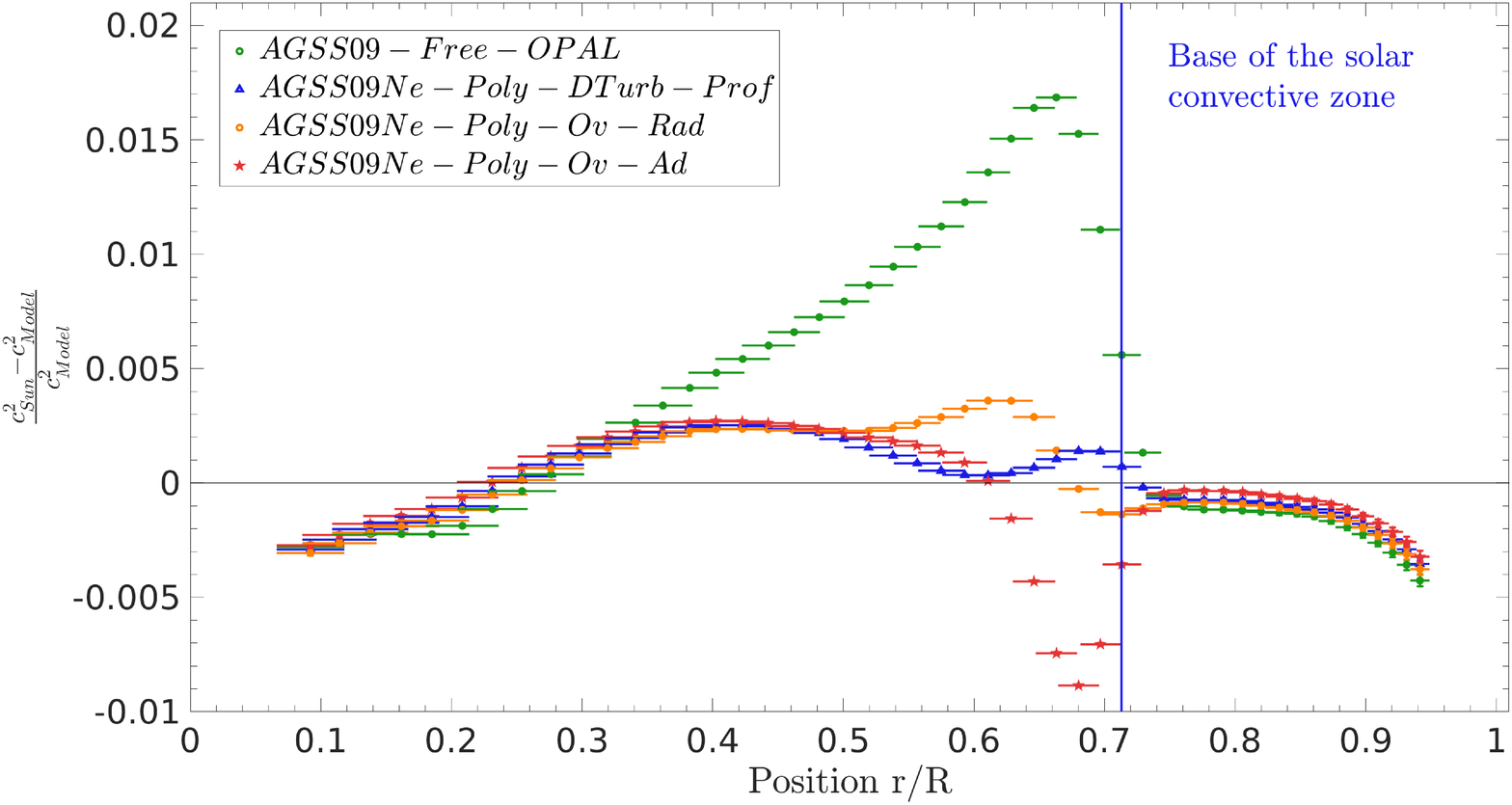}
	\caption{Relative differences in squared adiabatic sound speed between the Sun and solar models built using an ad-hoc modification of the mean Rosseland opacity and additional mixing of the chemical elements at the base of the convective zone.}
	\label{figC2PolyRad}
\end{figure*}

Here, we considered rather crude prescriptions for the extra-mixing at the base of the convective zone, using either instantaneous mixing or turbulent diffusion where diffusion coefficient was taken as a power law of density, following the prescription of \citet{Proffitt}. In the case of instantaneous mixing, we considered both the case of an adiabatic and radiative temperature gradient in the overshooting region. The results for the squared adiabatic sound speed are presented in Figure \ref{figC2PolyRad}, where we can see that adding an additional mixing in the form of turbulent diffusion provides a very good agreement near the base of the solar convective envelope. Instantaneous mixing only provides an improvement if the radiative temperature gradient is taken in the overshooting region. Setting the gradient to the adiabatic one induces the apparition of a large glitch feature in the sound speed profile, in strong disagreement with helioseismic data. 

\begin{figure*}
	\centering
	\includegraphics[width=0.75\linewidth]{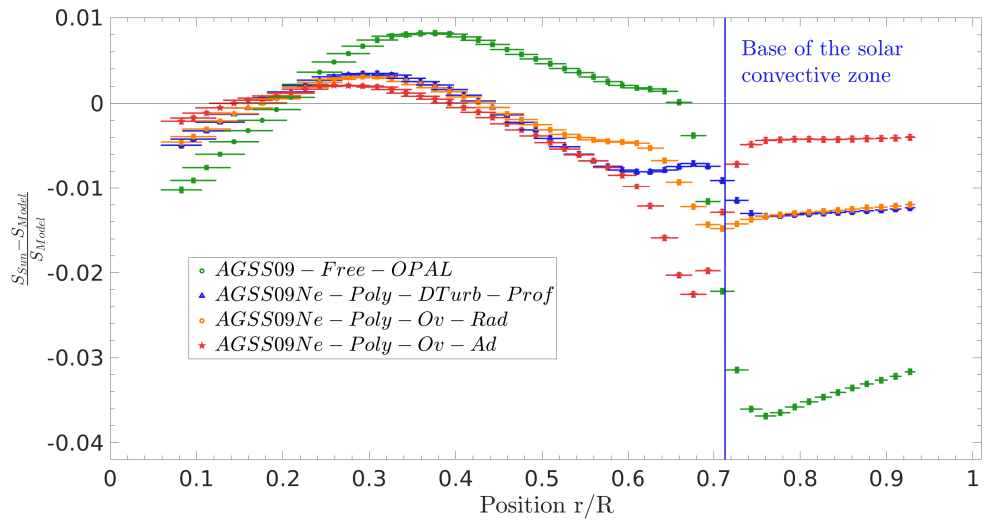}
	\caption{Relative differences in entropy proxy, $S_{5/3}$ between the Sun and solar models built using an ad-hoc modification of the mean Rosseland opacity and additional mixing of the chemical elements at the base of the convective zone.}
	\label{figSPolyRad}
\end{figure*}

As mentioned before, the sound speed inversion is actually insufficient to disentangle the various contributors to the solar modelling problem. This is illustrated in Figure \ref{figSPolyRad}, where we can see that adiabatic overshoot provides a much better agreement in terms of the height of the entropy proxy plateau in the solar convective zone than turbulent diffusion or overshoot with the radiative temperature gradient considered for the overshooting region. This actually illustrates the need for a different approach to modelling the transition in temperature gradient in the overshooting region and seems to advocate for transitions similar to those found in \citet{JCDOV}. In that sense, testing the impact of such transitions, in agreement with the glitch signal of the oscillation frequencies, on the entropy proxy inversion would provide very interesting insights on the potential degeneracies at play.

\begin{figure*}
	\centering
	\includegraphics[width=0.75\linewidth]{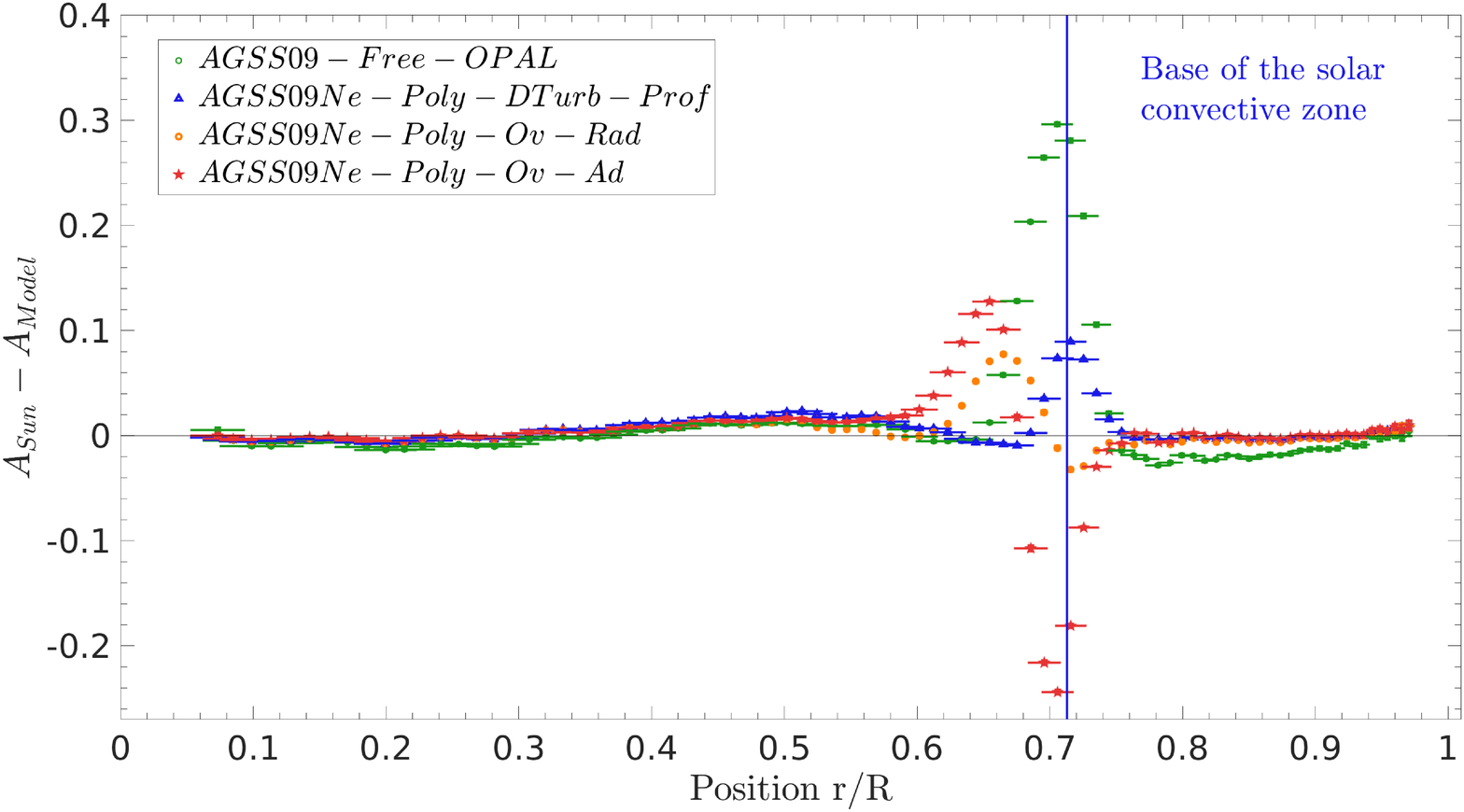}
	\caption{Differences in Ledoux discriminant, $A$, between the Sun and solar models built using an ad-hoc modification of the mean Rosseland opacity and additional mixing of the chemical elements at the base of the convective zone.}
	\label{figAPolyRad}
\end{figure*}

In addition to the inversions of the entropy proxy, we presented in \citet{BuldgenA} inversions of the so-called Ledoux discriminant, a proxy of the Brunt-Väisälä frequency, which allows to probe more local aspects of the temperature and chemical composition transition at the base of the convective zone. The results for the models including additional mixing are illustrated in Figure \ref{figAPolyRad} and results for a larger sample of models can be found in \citet{BULDGEN}. 

These inversion results allow us to gain more insight into the degeneracies of the solar modelling problem, as here we can see that each of the additional mixing prescriptions lead to different disagreements. For example, the adiabatic overshoot clearly leads to an inverted glitch signal and very large discrepancies, whereas both the turbulent diffusion and radiative overshoot show smaller discrepancies. Further testing the mixing prescriptions is required before concluding on the origins of the differences, as both thermal and chemical aspects are contained in the Ledoux discriminant profile. 

Taking into account the lithium and beryllium abundances as constraints for this extra-mixing might also prove crucial to be on more solid grounds to discuss the intensity and extension of the mixed region we include in modified solar models. 

From these investigations, we can indeed conclude that the solar modelling problem stems from various multiple small contributors, as multiple effects and their interplay may affect the agreement of the models with seismic inversions. While the issue is clearly complex and various fundamental ingredients are at play, we have shown that some of the degeneracies can be investigated by combining seismic inversions. 

\section{Conclusion and future prospects}

In this study, we have shown that no combination of the current standard ingredients for solar models could provide a satisfactory solution to the solar modelling problem. By combining the information of various seismic inversions, we have constrained the qualitative behaviour of the required opacity modification to bring models using the recent abundances of \citet{AGSS09} in better agreement with seismic data. The modification is largest at the base of the convective zone, in the temperature regime of an iron opacity peak and quickly drops to values of a few percents at higher temperatures. Moreover, we have found that additional chemical mixing beyond the convective envelope was also required. Both the opacity modification and the additional chemical mixing are physically motivated, as they are in agreement with the recent experimental measurements of \citet{Bailey} and the expectations of revised theoretical calculations (\citet{Zhao}, \citet{Pradhan}). It is worth noticing that none of the more recent opacity tables lead to an univoquous improvement of the situation.

Thus, it is clear that further opacity revisions might lead to an improvement of the current solar problem but other ingredients are certainly at play (see \citet{Ayukov2017} for a similar study and \citet{BULDGEN} for a more detailed description of the tests presented here), such as the equation of state, nuclear reaction rates, the hypotheses used for microscopic diffusion as well as the treatment of convective boundaries. Combining our structural inversions to glitch analyses of the transition in temperature gradient at the base of the convective envelope, as in \citet{JCDOV}, is the next step to improve our understanding of the Sun. 

In a more global perspective, the expected changes in the physical ingredients of solar models will impact our determinations of stellar fundamental parameters. Progress on this issue is paramount if we want to bring theoretical stellar models to a new level of accuracy in preparation for the upcoming PLATO mission and the exploitation of \textit{Kepler} and TESS data.

\bibliographystyle{phostproc}
\bibliography{BuldgenPoster.bib}

\end{document}